# HiTRACE-Web: an online tool for robust analysis of high-throughput capillary electrophoresis

Hanjoo Kim[1,2], Pablo Cordero[3], Rhiju Das[3,4*] and Sungroh Yoon[1,2*]

[1] Department of Electrical and Computer Engineering, Seoul National University, Seoul, Korea
[2] Bioinformatics Institute, Seoul National University, Seoul, Korea
[3] Program in Biomedical Informatics, Stanford University, Stanford CA 94305, USA
[4] Departments of Biochemistry and Physics, Stanford University, Stanford, CA 94305, USA

* To whom correspondence should be addressed: Tel: +1 650 723 5976; Email: rhiju@stanford.edu
* Correspondence may also be addressed to Sungroh Yoon: Tel: +82 2 880 1401; Email: sryoon@snu.ac.kr

**ABSTRACT**

To facilitate the analysis of large-scale high-throughput capillary electrophoresis data, we previously proposed a suite of efficient analysis software named HiTRACE (High Throughput Robust Analysis of Capillary Electrophoresis). HiTRACE has been used extensively for quantitating data from RNA and DNA structure mapping experiments, including mutate-and-map contact inference, chromatin footprinting, the EteRNA RNA design project and other high-throughput applications. However, HiTRACE is based on a suite of command-line MATLAB scripts that requires nontrivial efforts to learn, use, and extend. Here we present HiTRACE-Web, an online version of HiTRACE that includes standard features previously available in the command-line version as well as additional features such as automated band annotation and flexible adjustment of annotations, all via a user-friendly environment. By making use of parallelization, the on-line workflow is also faster than software implementations available to most users on their local computers. Free access: http://hitrace.org.

**INTRODUCTION**

Capillary electrophoresis (CE) is one of the most powerful and widely used nucleic acid separation techniques available (1). Its conventional application areas include genomic mapping, forensic identification, and genome sequencing (2,3). Recently combined with chemical probing methodologies, CE further provides a powerful and rapid means to map complex DNA and RNA structures at single-nucleotide resolution (4).

In chemical-probing-based RNA structure inference studies, a chemical reagent modifies the RNA of interest, either cleaving it or forming a covalent adduct with it. Commonly used reagents include hydroxyl radicals (5,6), dimethyl sulfate alkylation (DMS) (7), carbodiimide modification (CMCT) (8), and the SHAPE strategy using 2'-OH acylation (9). Subsequent reverse transcription detects the modification sites as stops to primer extension at nucleotide resolution. Traditionally, the resulting cDNA fragments were resolved in sequencing gels followed by individually quantifying band intensities. To resolve these fragments in a high-throughput fashion, capillary electrophoresis (CE) can be used. CE-based chemical probing can produce tens of thousands of individual electrophoretic bands from a single experiment, leading to recent breakthroughs in high-throughput nucleic acid structure mapping: e.g., automated inference of complex RNA structures (4,10-13) such as ribosomes (14), and viruses (15,16).

Analyzing a large number of electrophoretic traces from a high-throughput structure-mapping experiment is time-consuming and poses a significant informatics challenge. It requires a set of robust signal-processing algorithms for accurate quantification of the structural information embedded in the noisy traces. Current software methods for CE analysis include capillary automated footprinting analysis (CAFA) (10), ShapeFinder (11), high-throughput robust analysis for capillary electrophoresis (HiTRACE) (17), fast analysis of SHAPE traces (FAST) (18), and QuShape (19). The most recent of these methods have largely converged on the basic steps of analysis: preprocessing (such as selection of the data-containing range and baseline adjustment), deconvolution of co-loaded mapping signal traces and reference traces, alignment, peak detection, band annotation, peak fitting, signal decay and background subtraction.

Although these programs are all useful for semi-automated CE data analysis, they suffer from certain limitations. CAFA has effective peak fitting capabilities but lacks alignment and annotation features, thus necessitating laborious efforts for analyzing multiple capillaries. ShapeFinder and QuShape provide sophisticated signal alignment, annotation, and peak fitting capabilities, but have limited cross-capillary analysis capabilities, making these tools less efficient for analyzing data from large-scale experiments involving hundreds of capillaries. FAST is highly optimized for rapid band annotation but relies on the proprietary ABI utility programs, which has made it difficult to customize and extend to various experimental scenarios. HiTRACE is feature-rich and has been extensively used for high-throughput structure mapping studies such as the mutate-and-map strategy (13,20,21), chromatin footprinting, and the massively parallel RNA design project EteRNA (22). However, HiTRACE is a suite of command-line

MATLAB scripts that requires nontrivial efforts to learn, use, and extend, and a purchased license (The MathWorks, http://www.mathworks.com).

HiTRACE-Web is based on the HiTRACE workflow and thus inherits all of its core functionalities for high-throughput CE analysis. Going one step further, HiTRACE-Web provides an integrated and interactive on-line interface. In addition to the standard features previously available, HiTRACE-Web presents additional features such as automated band annotation and adjustment. By making use of powerful multi-core server processors, HiTRACE-Web also runs faster than the previous off-line implementations available to most users on their local computers. To the best of the authors' knowledge, HiTRACE-Web is the first on-line tool for high-throughput CE data analysis.

**METHOD OVERVIEW**

HiTRACE-Web takes as input a set of nucleic acid structure mapping profiles obtained from a number of capillaries. A profile represents the intensity of a nucleic acid sample in a capillary as a function of electrophoretic time. The peaks in a profile appear as bands in a gray-scale image. Band/peak locations represent individual residues of a nucleic acid sequence. Commonly used ABI sequencers (Life Technologies Corporation, Carlsbad, USA) can separate the products of hundreds of nucleotides, and we assume that each CE profile contains hundreds of bands. The final output of HiTRACE-Web is a set of aligned and annotated profiles with quantified band areas in text and image formats. To deliver the output, HiTRACE-Web works in multiple stages interleaved with user checkpoints: preprocessing, profile alignment, band annotation, peak fitting and quantification. Figure 1 shows the flowchart of the analysis pipeline.

In preprocessing, the user needs to provide information on capillary channel compositions. To facilitate the analysis process, it is common in typical CE experiments to co-load samples with a reference ladder which fluoresces in a different color. Multiple spectral channels thus exist in each capillary, and the user should specify which channels contain the structure mapping signal and the reference (typically a ladder) (Step A.1).

The next step in preprocessing is to select the range of analysis, since typical profiles carry information only within a subset of the entire electrophoresis run. HiTRACE-Web utilizes edge-detection techniques (23) to select proper regions for further analysis automatically (Step A.2). It is also possible for the user to set the region manually. HiTRACE-Web then carries out a series of additional preprocessing operations on the selected region of capillaries such as adjustment of constant baseline (Step A.3). More details of the preprocessing operations can be found in (17).

The second stage in HiTRACE-Web is alignment. Profiles need to be aligned to each other because the products in different capillaries are subject to slightly different electrophoretic conditions and their detection times are then shifted and scaled compared to each other. Due to the finite number of capillaries an ABI sequencer can handle at a time (e.g., 96 for ABI3730), CE experiments using hundreds of capillaries consist of multiple batches. We designate the first capillary in each batch as the reference and use it for within-batch alignment. There, we align each profile to the reference by finding the optimal shift and scaling amounts that maximize its correlation to the reference (part of Step B.1). After the within-batch alignment, we carry out additional alignment procedures: between-batch adjustment for global alignment (part of Step B.1), adjustment of smooth baseline (Step B.2), and dynamic-programming-based piece-wise-linear alignment for fine-tuning local disagreements (17) (Step B.3).

After alignment, the user checks intermediate results (Step C.1) and specifies the sequence probed and the type of chemical modifications applied to each capillary (Step C.2). The user can currently select among nine options: three types of chemical reagents (DMS, CMCT, and SHAPE), four types of dideoxynucleotides (ddGTP, ddATP, ddTTP, and ddCTP), no modification, or other. HiTRACE-Web provides a convenient and intuitive graphical interface for selecting signal and reference channels and specifying capillary modification types. This specification provides HiTRACE-Web with information on where bands should appear in the data, e.g., primarily at A and C positions for DMS modification of RNA data (24), and with annotations carried forward to the final result files (see below).

Following the profile alignment is the band annotation procedure (Step D.1), which refers to the process of mapping each band in an electrophoretic profile to a position in the nucleic acid sequence. For verification, visual inspection of band annotation results is normally inevitable to certain extent, but the manual band annotation step takes significant human efforts when there are a large number of profiles. To address this issue, HiTRACE-Web provides an automated band annotation functionality, which allows users to complete initial assignments of hundreds of bands in hundreds of profiles in the order of seconds. More details of the automated band assignment algorithm are beyond the scope of this server-focused paper and will be described elsewhere. HiTRACE-Web also provides an interactive interface to adjust the band annotation manually so that users can correct any suboptimal assignment they find.

In the last stage, HiTRACE-Web performs peak fitting (Step D.2) to approximate a profile as a sum of Gaussian curves, each of which is centered at the intensity peak location. The peak amplitudes are selected in the least-squares sense by using a standard optimization technique, thus minimizing the deviations of the Gaussian model from the CE profiles (17). As the final output, HiTRACE-Web reports the peak quantification results including the area and location of each band and exports data annotations, sequence, and structure to an RNA data (RDAT) formatted file (25). The RDAT file can then be submitted to the RNA Mapping Database (RMDB) repository for data sharing (http://rmdb.stanford.edu/submit) or structure server for secondary structure model estimation (http://rrmdb.stanford.edu/structureserver) (25), and links to these tools are provided.

HiTRACE-web leaves correction for attenuation of band intensity for longer products ('signal decay'), background subtraction, and final normalization to the user. These post-processing tasks are carried out differently by independent groups who have made distinct *ad hoc* assumptions (19,26,27), and indeed these tasks are left out of some applications, such as the mutate-and-map technique (21), due to the introduction of noise. Robust experimental and computational approaches for post-processing chemical mapping data are in development (T. Mann, PC, RD, in prep.), as are additional features including secondary structure display and error estimation in automatic sequence assignment. Inclusion into HiTRACE-Web will allow these advances to be disseminated rapidly to the RNA structure mapping community.

**WEB SERVER**

The HiTRACE-Web server utilizes the Apache HTTP Server (The Apache Software Foundation, http://httpd.apache.org/) for basic web services and the MySQL Server (Oracle Corporation, http://www.mysql.com/) for internal data management. For client-side programming, we used CodeIgniter (EllisLab, http://ellislab.com/codeigniter), an open-source web application framework, and coded the web pages in PHP (The PhP Group, http://php.net/) and JavaScript (Mozilla Foundation, https://developer.mozilla.org/en/docs/JavaScript) with jQuery (The jQuery Foundation, http://jquery.com/) and jQuery user-interface (UI) plugins. Interactive UI components also use the canvas elements defined in the HTML5 standard (World Wide Web Consortium, http://www.w3.org/TR/html5/). We used Ajax (http://en.wikipedia.org/wiki/Ajax_(programming)) for asynchronous data transfers between the client and server sides. On the server side, we used Gearman (http://gearman.org), an open-source application framework, to schedule multiple requests. Each user request is handled by a worker written in PHP. The worker creates template code in MATLAB that contains input parameters and links to the user data. The worker also forks the MATLAB interpreter so that it can execute the template code. Most of the time-consuming operations are multi-threaded, and the current HiTRACE-Web server runs on a 48-core machine (four on-board AMD Opteron 6172 processors with 256GB main memory; Ubuntu Linux version 3.2.0-29). In this environment, processing 52 capillaries (with 60 bands per capillary) takes a few minutes including manual adjustments. HiTRACE-Web supports most of the widely used web browsers including Google Chrome (Version 25.0 or later; http://www.google.com/chrome/), Mozilla Firefox (Version 19.0 or later; http:// www.mozilla.org/), Apple Safari (Version 6.0 or later; http:// www.apple.com/safari/), and Microsoft Internet Explorer (Version 9.0 or later; http://windows.microsoft.com).

Figure 2 explains the overall flow of CE data analysis using HiTRACE-Web. The input file is a zip-compressed collection of .ab1 or .fsa files from ABI sequencers in the ABIF file format. If there are multiple batches, each batch should be presented in a subfolder, as is the typical output from ABI sequencers. After uploading the input file, the user needs to specify the signal and reference channels. HiTRACE-Web provides a graphical interface for channel selection (Figure 2A). Next, the region of interest is specified either manually or automatically (Figure 2B). HiTRACE-Web then carries out the alignment of the profiles in the selected region, following the procedures outlined previously. The user can visually inspect the intermediate result after alignment (Figure 2C) and go over the previous stages if needed. If satisfactory, the user can start specifying the type of chemical modification data present in each capillary using the graphical interface HiTRACE-Web provides (Figure 2D). The user can then perform band annotation (Figure 2E). HiTRACE-Web permits either manual or automated band annotation; in practice, both types of annotations complement each other: performing automated band annotation first and then adjusting the result manually allows a large number of bands to be annotated quickly and robustly. After band annotation, HiTRACE-Web carries out peak fitting and quantification. The results are provided as downloadable images (Figure 2F), tab-delimited text files (with quantified areas; one column per capillary and one row per residue), and an RDAT file that can be submitted to the repository and structure modeling server available at the RNA Mapping Database (25). More detailed instructions to each analysis stage and explanations of results are available at the HiTRACE-Web homepage. A complete tutorial with example data is also available, with specific help at each step.

Figure 3A and 3B show the correlation of quantification results between HiTRACE-Web and HiTRACE (17) for two different users. The high level of Pearson's correlation coefficients ($R^2$ of 0.9996 and 0.9999; $P<0.001$) between band intensities quantified with HiTRACE-Web and those quantified with HiTRACE indicates lack of any major systematic deviations induced by HiTRACE-Web. To test the consistency in band quantification between different users, we let two independent users carry out quantification of the same data using HiTRACE-Web and HiTRACE, as shown in Figure 3C and 3D, respectively. Both tools resulted in excellent agreement between the independent analyses ($R^2$ of 0.9993 and 0.9986; $P<0.001$). The data used is from an RNA structure mapping study using the mutate-and-map strategy (13,20,21). The mapped sequence length was 92 nucleotides, and the data describes six different mapping experiments including SHAPE, DMS and reference ladders. The total number of bands was 552.

**SUMMARY**

We developed HiTRACE-Web (freely available at http://hitrace.org), an online server for rapid and robust analysis of large collections of profiles obtained from high-throughput CE experiments. Recent generations of high-throughput DNA and RNA structure mapping studies give hundreds of CE profiles each containing hundreds of bands. To isolate signals from such a large collection of profiles, we previously developed a signal-processing pipeline that consists of multiple stages including preprocessing, alignment, annotation, and band quantification. HiTRACE-Web now enables users to follow the whole pipeline through a user-friendly, integrative, and interactive web environment. It is our hope that HiTRACE-Web can contribute to large-scale structure inference studies based on chemical probing and CE separation by providing an effective and easily accessible data analysis framework.

**ACKNOWLEDGEMENT**

The authors thank members of the Das lab and the Yoon lab for extensive testing of the web server.

**FUNDING**

This work was supported by the National Research Foundation of Korea funded by the Ministry of Education, Science and Technology [No. 2011-0009963 and No. 2012-0008475 to S.Y. in part]; Samsung Electronics Co., Ltd. [No. 0423-20130021 to S.Y. in part]; a CONACyT pre-doctoral scholarship [to P.C. in part]; Burroughs-Wellcome Foundation Career Award at the Scientific Interface [to R.D. for computational work in part]; and the National Institutes of Health [R01 GM102519 to R.D. in part]. Funding for open access charge: Seoul National University.

**REFERENCES**

1. Weinberger, R. (2000) Practical capillary electrophoresis. Second ed. Academic Press San Diego, CA:.
2. Luckey, J.A., Drossman, H., Kostichka, A.J., Mead, D.A., D'Cunha, J., Norris, T.B. and Smith, L.M. (1990) High speed DNA sequencing by capillary electrophoresis. Nucleic acids research, **18**, 4417-4421.
3. Woolley, A.T. and Mathies, R.A. (1995) Ultra-high-speed DNA sequencing using capillary electrophoresis chips. Analytical chemistry, **67**, 3676-3680.
4. Weeks, K.M. (2010) Advances in RNA structure analysis by chemical probing. Current opinion in structural biology, **20**, 295-304.
5. Das, R., Kudaravalli, M., Jonikas, M., Laederach, A., Fong, R., Schwans, J.P., Baker, D., Piccirilli, J.A., Altman, R.B. and Herschlag, D. (2008) Structural inference of native and partially folded RNA by high-throughput contact mapping. Proceedings of the National Academy of Sciences of the United States of America, **105**, 4144-4149.
6. Kim, J., Yu, S., Shim, B., Kim, H., Min, H., Chung, E.Y., Das, R. and Yoon, S. (2009) A robust peak detection method for RNA structure inference by high-throughput contact mapping. Bioinformatics, **25**, 1137-1144.
7. Tijerina, P., Mohr, S. and Russell, R. (2007) DMS footprinting of structured RNAs and RNA-protein complexes. Nature protocols, **2**, 2608-2623.
8. Walczak, R., Westhof, E., Carbon, P. and Krol, A. (1996) A novel RNA structural motif in the selenocysteine insertion element of eukaryotic selenoprotein mRNAs. Rna, **2**, 367-379.
9. Merino, E.J., Wilkinson, K.A., Coughlan, J.L. and Weeks, K.M. (2005) RNA structure analysis at single nucleotide resolution by selective 2'-hydroxyl acylation and primer extension (SHAPE). Journal of the American Chemical Society, **127**, 4223-4231.
10. Mitra, S., Shcherbakova, I.V., Altman, R.B., Brenowitz, M. and Laederach, A. (2008) High-throughput single-nucleotide structural mapping by capillary automated footprinting analysis. Nucleic acids research, **36**, e63.
11. Vasa, S.M., Guex, N., Wilkinson, K.A., Weeks, K.M. and Giddings, M.C. (2008) ShapeFinder: a software system for high-throughput quantitative analysis of nucleic acid reactivity information resolved by capillary electrophoresis. Rna, **14**, 1979-1990.

12. Das, R., Karanicolas, J. and Baker, D. (2010) Atomic accuracy in predicting and designing noncanonical RNA structure. Nature methods, **7**, 291-294.
13. Kladwang, W. and Das, R. (2010) A mutate-and-map strategy for inferring base pairs in structured nucleic acids: proof of concept on a DNA/RNA helix. Biochemistry, **49**, 7414-7416.
14. Deigan, K.E., Li, T.W., Mathews, D.H. and Weeks, K.M. (2009) Accurate SHAPE-directed RNA structure determination. Proceedings of the National Academy of Sciences of the United States of America, **106**, 97-102.
15. Wilkinson, K.A., Gorelick, R.J., Vasa, S.M., Guex, N., Rein, A., Mathews, D.H., Giddings, M.C. and Weeks, K.M. (2008) High-throughput SHAPE analysis reveals structures in HIV-1 genomic RNA strongly conserved across distinct biological states. PLoS biology, **6**, e96.
16. Watts, J.M., Dang, K.K., Gorelick, R.J., Leonard, C.W., Bess, J.W., Jr., Swanstrom, R., Burch, C.L. and Weeks, K.M. (2009) Architecture and secondary structure of an entire HIV-1 RNA genome. Nature, **460**, 711-716.
17. Yoon, S., Kim, J., Hum, J., Kim, H., Park, S., Kladwang, W. and Das, R. (2011) HiTRACE: high-throughput robust analysis for capillary electrophoresis. Bioinformatics, **27**, 1798-1805.
18. Pang, P.S., Elazar, M., Pham, E.A. and Glenn, J.S. (2011) Simplified RNA secondary structure mapping by automation of SHAPE data analysis. Nucleic acids research, **39**, e151.
19. Karabiber, F., McGinnis, J.L., Favorov, O.V. and Weeks, K.M. (2013) QuShape: rapid, accurate, and best-practices quantification of nucleic acid probing information, resolved by capillary electrophoresis. Rna, **19**, 63-73.
20. Kladwang, W., Cordero, P. and Das, R. (2011) A mutate-and-map strategy accurately infers the base pairs of a 35-nucleotide model RNA. Rna, **17**, 522-534.
21. Kladwang, W., VanLang, C.C., Cordero, P. and Das, R. (2011) A two-dimensional mutate-and-map strategy for non-coding RNA structure. Nature chemistry, **3**, 954-962.
22. Bida, J.P. and Das, R. (2012) Squaring theory with practice in RNA design. Current opinion in structural biology, **22**, 457-466.
23. Canny, J. (1986) A computational approach to edge detection. Pattern Analysis and Machine Intelligence, IEEE Transactions on, 679-698.
24. Cordero, P., Kladwang, W., VanLang, C.C. and Das, R. (2012) Quantitative dimethyl sulfate mapping for automated RNA secondary structure inference. Biochemistry, **51**, 7037-7039.
25. Cordero, P., Lucks, J.B. and Das, R. (2012) An RNA Mapping DataBase for curating RNA structure mapping experiments. Bioinformatics, **28**, 3006-3008.
26. Kladwang, W., VanLang, C.C., Cordero, P. and Das, R. (2011) Understanding the errors of SHAPE-directed RNA structure modeling. Biochemistry, **50**, 8049-8056.
27. Aviran, S., Lucks, J.B. and Pachter, L. (2011) RNA structure characterization from chemical mapping experiments. Arxiv preprint arXiv:1106.5061.

**FIGURES LEGENDS**

Figure 1. Flowchart of HiTRACE-Web analysis pipeline. Each of the steps marked with an asterisk corresponds to a tab in the web server implementation. The first stage is preprocessing in which the user can confirm or refine the signal and the reference channels, proper data regions for analysis, and the correction of constant baselines. HiTRACE-Web carries out profile alignment in the second stage. Both linear (for within-batch and between-batch alignment) and piece-wise-linear alignment is performed. After alignment, the user can check the intermediate result, provide sequence and structure information, and specify types of chemical reagents in the third stage. In the next stage, HiTRACE-Web provides an automated band annotation functionality, which allows users to complete initial assignments of hundreds of bands in hundreds of profiles in the order of seconds. HiTRACE-Web then performs peak fitting to approximate a profile as a sum of Gaussian curves.

Figure 2. Overview of data analysis using HiTRACE-Web. (A) Users can select the reference and signal channels using the graphical interface. (B) HiTRACE-Web provides a feature to select the valid region of interest automatically. Alternatively, users can set the range manually. The red rectangles in the figure represent the regions of interest. (C) Before advancing to the next step, HiTRACE-Web provides a snapshot of intermediate results, so that the user can go over the previous steps if needed. (D) HiTRACE-Web provides an intuitive graphical interface to specify the chemical modifications made to capillaries. For each of them, the user can select among nine color-coded options: three chemical reagents (dimethyl sulfate alkylation (DMS) (7), carbodiimide modification (CMCT) (8), the SHAPE strategy using 2'-OH acylation (9)), reference ladders using four dideoxynucleotides (ddGTP, ddATP, ddTTP, and ddCTP), no modification, and other. (E) HiTRACE-Web can carry out band annotation in an automated fashion, thus reducing the analysis time substantially. Furthermore, HiTRACE-Web provides a user-friendly interface for users to fine-tune band annotation results manually. Red circles represent the residue locations. Each type of nucleotide is associated with a different color (G: green, C: cyan, U: blue, and A: red). By clicking the image, user can (de)select the position of individual residues. The auto-assigned bands are shown on the right side in gray for easier referencing when manually adjusting the assignment. (F) HiTRACE-Web reports a set of aligned and annotated profiles with quantified peak areas in the image, tab-delimited text, and RDAT (25) formats users can download for further uses.

Figure 3. Confirming consistency in band quantification results between tools and analyses. (A-B) Correlation of quantification results between HiTRACE-Web and HiTRACE (17) for two independent users. These plots indicate lack of any major systematic deviations induced by HiTRACE-Web. (C-D) HiTRACE-Web gives the same level of consistency between independent analyses as HiTRACE. The data used is from an RNA structure mapping study using the mutate-and-map strategy (13,20,21). A 92-nucleotide RNA sequence was treated in six different mapping conditions including SHAPE, DMS and ddTTP, giving 552 bands in total.

# Figure 1

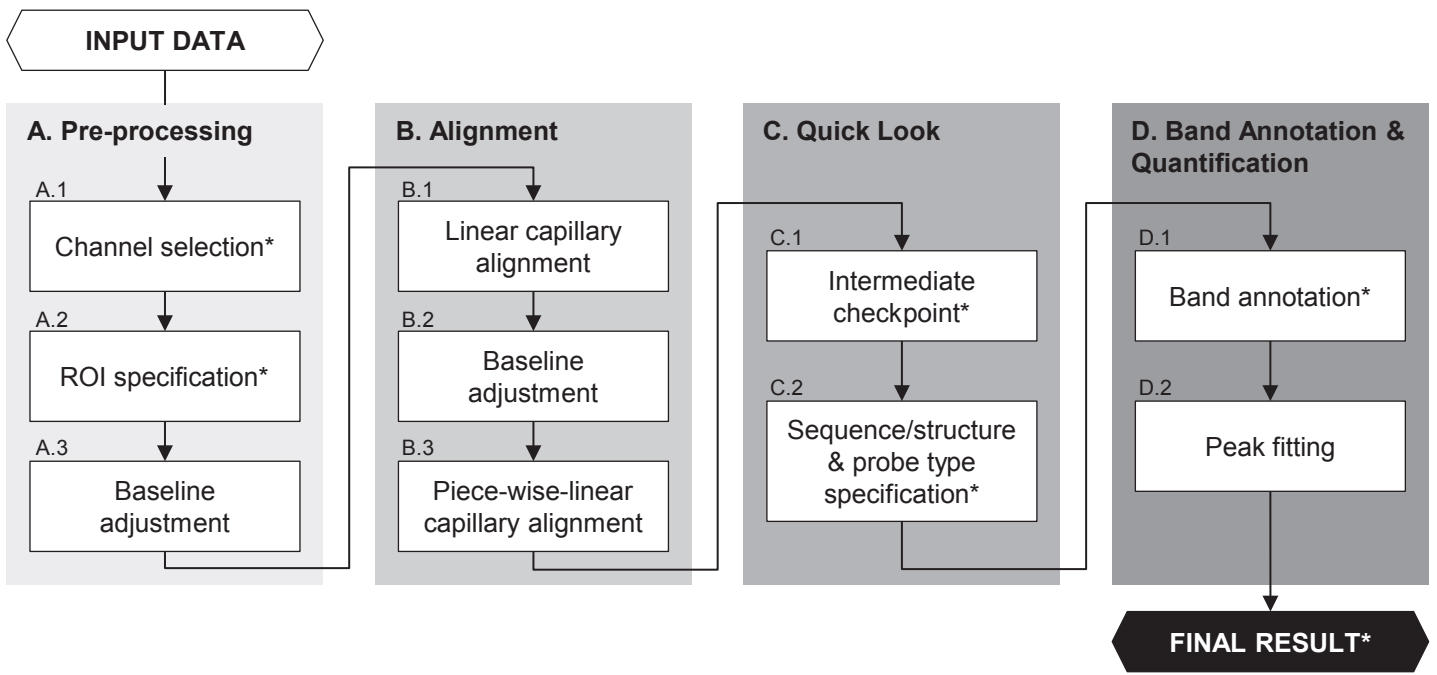

**Figure 2**

# Figure 3

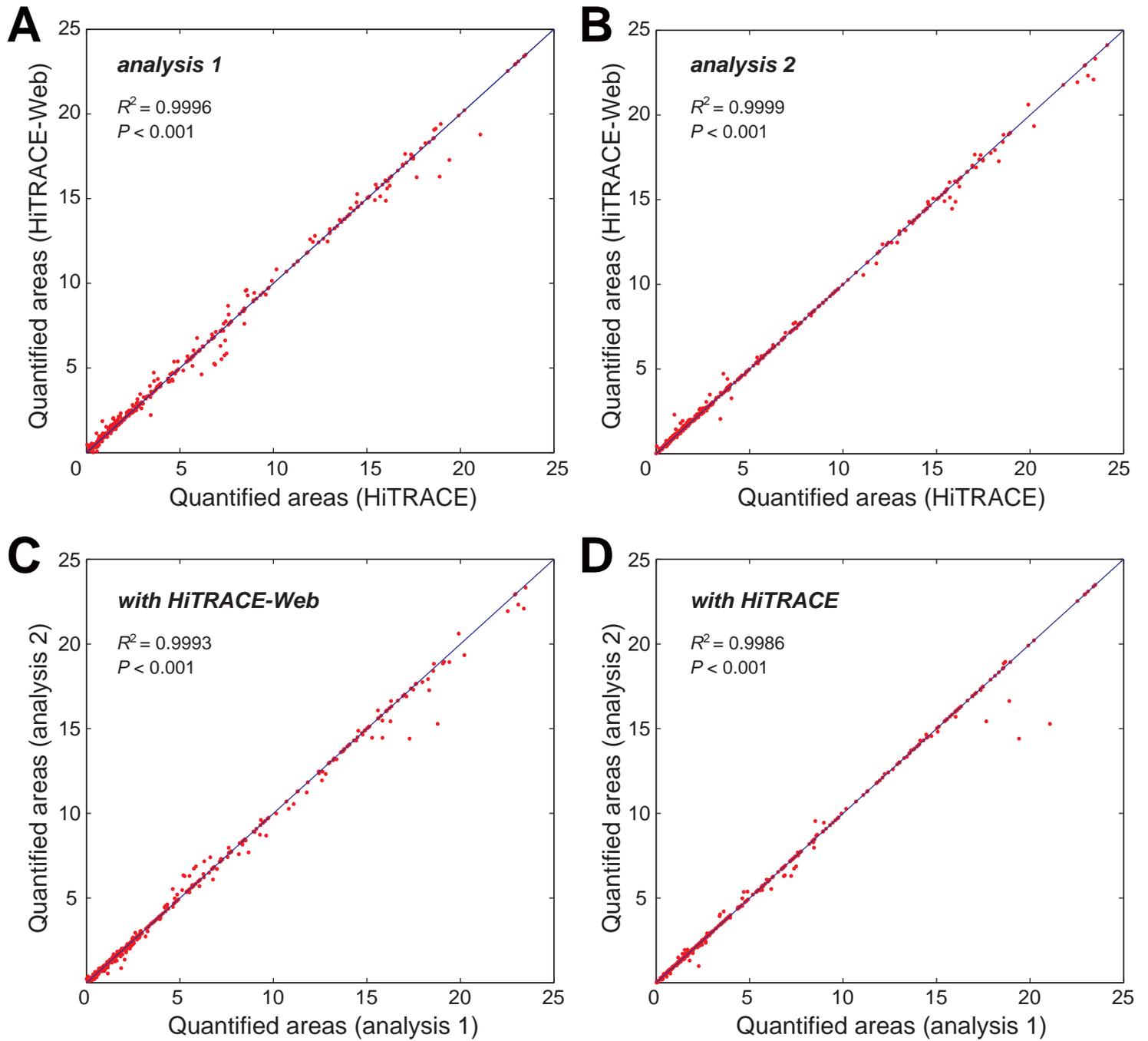